\documentclass[twocolumns,a4paper]{aa}
\usepackage{amsmath} 
\usepackage{relsize}                                  
\usepackage{epsfig}
\usepackage{natbib} 
\bibpunct{(}{)}{;}{a}{}{,} 
\usepackage{graphicx}
\begin{document}
\title{Spatial deconvolution of spectropolarimetric data: an application to quiet Sun magnetic elements}
\author{C. Quintero Noda\inst{1,2} \and A. Asensio Ramos\inst{1,2} \and D. 
Orozco Su\'arez\inst{1,2} \and B. Ruiz Cobo\inst{1,2}}
\institute{Instituto de Astrof\'isica de Canarias, E-38200, La Laguna, Tenerife, 
Spain.\ \email{cqn@iac.es} \and
Departamento de Astrof\'isica, Univ. de La Laguna, La Laguna, Tenerife, E-38205, 
Spain}

\date{Received/Accepted}

\abstract 
{One of the difficulties in extracting reliable information about the 
thermodynamical and magnetic properties of solar plasmas from 
spectropolarimetric observations is the presence of light dispersed inside the 
instruments, known as stray light.}
{We aim to analyze quiet Sun observations after the spatial deconvolution of the 
data. We examine the validity of the deconvolution process with noisy 
data as we analyze the physical properties of quiet Sun magnetic elements.}
{We used a regularization method that decouples the Stokes 
inversion from the deconvolution process, so that large maps can be quickly 
inverted without much additional computational burden. We applied the method on 
Hinode quiet Sun spectropolarimetric data. We examined the spatial and 
polarimetric properties of the deconvolved profiles, comparing them with the 
original data. After that, we inverted the Stokes profiles using the Stokes 
Inversion based on Response functions (SIR) code, which allow us to obtain the 
optical depth dependence of the atmospheric physical parameters.}
{The deconvolution process increases the contrast of continuum images and makes the magnetic structures sharper. The deconvolved Stokes $I$ profiles reveal the presence of the Zeeman splitting while the Stokes $V$ profiles significantly change their amplitude. The area and amplitude asymmetries of these profiles increase in absolute value after the deconvolution process. We inverted the original Stokes profiles from a magnetic element and found that the magnetic field intensity reproduces the overall behavior of theoretical magnetic flux tubes, that is, the magnetic field lines are vertical in the center of the structure and start to fan when we move far away from the center of the magnetic element. The magnetic field vector inferred from the deconvolved Stokes profiles also mimic a magnetic flux tube but in this case we found stronger field strengths and the gradients along the line-of-sight are larger for the magnetic field intensity and for its inclination. Moreover, the discontinuity between the magnetic and non magnetic environment in the flux tube gets sharper.
}
{The deconvolution process used in this paper reveals information that the 
smearing induced by the point spread function (PSF) of the telescope hides. Additionally, the deconvolution is done with a low computational load, making it appealing for its use on the analysis of large data sets.}
\titlerunning{Spatial deconvolution of spectropolarimetric data}
\authorrunning{Quintero Noda et al.}
\maketitle

\section{Introduction}

Observations of the Sun from the Earth are always limited by the presence of the atmosphere, which strongly disturbs the images. A solution to this problem is to place the telescopes in space satellites, which produce observations without any (or limited) atmospheric aberrations. Recent examples of these atmospheric-free observations are the Hinode mission \citep{Kosugi2007}, especially the spectropolarimeter \citep[SP,][]{Lites2013} of the solar optical telescope \citep[SOT,][]{Tsuneta2008}, and the vector magnetograph IMaX \citep{MartinezPillet2011} on board the Sunrise balloon \citep{Solanki2010}. Although the images from space are not affected by atmospheric seeing, the optical properties of the instruments still limit the observations. In the case of diffraction limited observations, the point spread function (PSF) establishes the maximum allowed spatial resolution, defined as the distance between two nearby structures that can be properly distinguished. In space observations, the central core of the PSF is typically dominated by the Airy disk, which is a consequence of a physical limitation imposed by the diffraction. Even in a diffraction limited instrument, real PSFs have typically the shape of the Airy pattern, with very extended tails. These tails do not limit the spatial resolution but induce a dispersion of the light from different parts of the image, leading to what is commonly termed as stray light or dispersed light. This effect produces that light observed in a spatial location at the focal plane is a combination of the light emitted in the object at relatively distant spatial locations. Therefore, the contrast of the object (defined as the pixel-to-pixel variation of the illumination normalized to the average illumination) measured in the focal plane is typically smaller than the contrast in the original object.

The presence of stray light is important both for imaging instruments and slit spectropolarimeters. A first successful attempt to correct for this effect in imaging instruments was carried out by \cite{MartinezPillet1992}, where an analytical PSF with long tails was proposed and the image was deconvolved from it following a least-squares approach. Another method could be to consider the stray light contamination as the sum of two components: a spectrally dispersed component and a parasitic component of the spectrally undispersed light caused by scattering inside the spectrograph \citep{Beck2011}. In addition, we can also find in the literature the multi-object multi-frame blind deconvolution technique \citep{vanNoort2005} which has been used to correct for all the perturbing effects of the atmosphere and the instrument. The novelty of the latter method is that a very general functional form for the PSF is proposed and blindly estimated from the observations, together with the corrected images.

The case of slit spectrographs is more complicated because the images are not immediately available. Instead, an image is constructed by adding the different scanning steps of the slit at different times. Therefore, one has to make some assumptions about the stability of the object so that the reconstructed images can be used for deconvolution. Furthermore, given that the spectral resolution of slit spectropolarimeters is much larger than that of image instruments, the number of monochromatic images is much larger. This makes the application of any deconvolution scheme a much more computationally heavy task. For this reason, it has been customary to postpone the treatment of dispersed light to the inversion phase of the spectropolarimetric data. A physical model (atmospheric model$+$radiative transfer) is proposed to explain the observed Stokes profiles, and an ad-hoc contamination is linearly added to account for the stray light. It is possible to find different ways of computing this contamination in the literature: from local approaches that compute the stray light in a box of $N \times N$ pixels around the pixel of interest \citep{Orozco2007b} to global approaches that use an average Stokes $I$ profile in the whole field-of-view. Global approaches are preferred to local ones for different reasons \citep{Asensio2011}, essentially because the use of local approaches somehow make the inversion process uncontrollable.

One way to proceed when carrying out an inversion of spectropolarimetric data is to simultaneously do the inversion and the image deconvolution. The first effort in this direction has been carried out by \cite{vanNoort2012}, in which a standard inversion code for the Stokes parameters \citep{frutiger00} is modified to simultaneously take the presence of the spatial coupling induced by the PSF into account. This represents the first realistic approach to a full inversion of the Stokes profiles without an ad-hoc treatment of the stray light. The approach followed by \cite{vanNoort2012} is computationally complex. The reason is that the inversion of the Stokes profiles is carried out simultaneously with the spatial deconvolution, using a Levenberg-Marquardt algorithm, without a distinction between the two processes. This algorithm needs to compute and invert a Hessian matrix that is very large. To minimize the computational load, \cite{RuizCobo2013} used a simplified approach in which the inversion is carried out in two steps: in the first step the spectropolarimetric data is deconvolved from the known PSF using a regularization based on a Karhunen-Lo\`{e}ve transformation or principal component analysis PCA \citep[see,][]{Loeve1955} and then inverted using standard Stokes inversion codes. In this paper, we explain the technique \citep[only briefly presented in][]{RuizCobo2013} in detail, and use it to analyze spectropolarimetric observations of a quiet Sun magnetic element obtained with Hinode/SP.

The magnetic field of the solar surface is structured in a wide range of scales. The largest are the sunspots and can reach sizes of Mm. If we move to smaller structures we find pores, plages, faculae, or small magnetic elements that could reach a size of 100 km. Early spectropolarimetric observations have revealed many of the fundamental properties of these structures, finding, for example, that the magnetic field intensity of these magnetic elements should be of the order of kG values \cite[see, for instance,][]{Stenflo1973}. Combined with basic magnetohydrodynamical (MHD) theory, this deduction led to the development of the thin flux tube model \citep{Steiner1998,Spruit1979,Parker1976} as a fundamental element on the structure of the photosphere. As time passes, the designs of solar telescopes has improved, providing new observational data to analyze their apparent size, brightness, field structure, dynamics, and evolution. Works like \cite{MartinezGonzalez2012,Laag2010,Viticchie2010,Rezaei2007,Berger2004,Dominguez2003,Berger2001,%
vanBallegooijen1998,Berger1996,Keller1992,Muller1983}, and the reviews of \cite{Solanki2006} and \cite{deWijn2009} roughly provide a complete picture of our current knowledge. All of these studies generally support the idea that virtually all of the small-scale structure in active and quiet network regions is composed of filamentary flux tubes of kG magnetic field strength.
 
In the present work, we explain in detail  the spatial deconvolution technique employed for the first time in \cite{RuizCobo2013}. In addition, we also apply this method to quiet Sun Hinode/SP data for the first time, aiming to take advantage of the spatial deconvolution process to analyze the physical properties of quiet Sun magnetic elements. 

\section{Observations and data analysis}
\label{observations}

\subsection{Observations}

\begin{figure*}[t]
\vspace{-1.5cm}
\centering
\includegraphics[width=17cm]{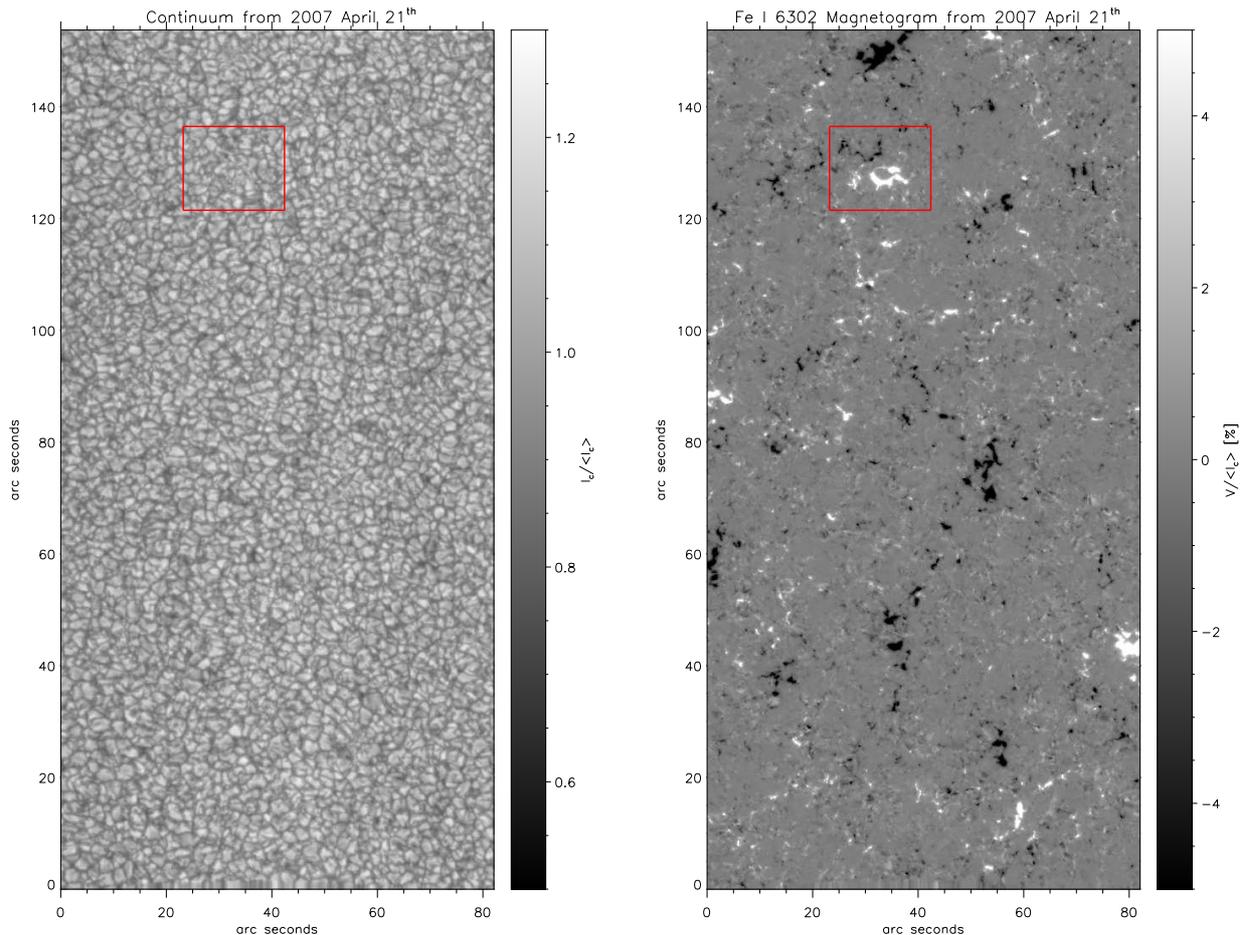}
\caption{Continuum map, left, and Fe~{\sc i} 6302.5 \AA \ magnetogram, right. The presence of small and not very common magnetic patches indicates that this map corresponds to a very quiet Sun region. Red square marks the magnetic element studied in detail.}
\label{map}
\end{figure*}

The polarimetric data we used were acquired with the spectropolarimeter \citep[SP;][]{Lites2013} on board the Hinode spacecraft \citep{Kosugi2007}. We selected a data set with a field of view of $82^{\prime\prime}\times164^{\prime\prime}$ recorded at disk center on April 21$^{\rm th}$, 2007 (see Figure \ref{map}). The SP instrument measures the Stokes vector of the Fe~{\sc i} line-pair at 630 nm with a spectral and spatial sampling of 2.15 pm pixel$^{-1}$ and 0.16$^{\prime\prime}$, respectively. The exposure time is 12.8 s per slit position, making it possible to achieve a noise level of $7.0\times10^{-4}~I_{c}$ in Stokes $V$ and $7.2\times10^{-4}~I_{c}$ in Stokes $Q$ and $U$. Here $I_{c}$ refers to the mean continuum intensity in the granulation. This data set displays a signal-to-noise ratio $\sqrt{12.8/4.8}$ times higher than that of the \textit{normal} Hinode/SP maps in which the exposure time is 4.8 s. To calibrate the spectra, we averaged the intensity profile from the whole map and compared it with the Fourier transform spectrometer spectral atlas \citep{Kurucz1984,Brault1987} once it was convolved with the spectral PSF of Hinode. A similar calibration was done by \cite{CabreraSolana2007} (see, Eq. 1). We found a difference between the intensity of both profiles which can be interpreted as parasitic light inside the instrument, defined as ``veil'' in \cite{RuizCobo2013}. In the present observation, the estimated value for this veil, is $C$=0.0357 referred to the continuum signal $I_c$. The data is finally corrected as $I^{final}(\lambda)=(I^{or}(\lambda)-C)/(I_c-C)$. We subtracted this value from the continuum intensity before normalization. Since we aim to analyze strong magnetic elements in the quiet Sun (see black and white structures in Figure \ref{map}), we selected one isolated magnetic structure, far from the edges of the map (see the red square), with strong longitudinal field signals, to study its properties in detail.

\subsection{Deconvolution}

We face the problem of correcting two-dimensional spectropolarimetric data from the perturbation introduced by the PSF of the Hinode solar optical telescope. We obtained the two dimensional data by scanning a slit on the surface of the Sun and recording the information of the four Stokes profiles $(I,Q,U,V)$ on each point along the slit for a set of discrete wavelength points around the 630 nm Fe~{\sc i} doublet. As a consequence, the data can be considered to be four three-dimensional cubes of images. We use the notation $\mathbf{I}(\lambda)$, $\mathbf{Q}(\lambda)$, $\mathbf{U}(\lambda)$, and $\mathbf{V}(\lambda)$ to refer to observed images at a certain wavelength $\lambda$. In practice, given the scanning process, these are not strictly speaking images because each column of the image is taken at a different time.

In general, in the standard image formation paradigm, the observed image \textbf{I} (for simplicity we focus on Stokes $I$, but the same expressions apply to any Stokes parameter given the linear character of the convolution operator) that one obtains in the detector after degradation by the atmosphere and the optical devices of the telescope at a given wavelength can be written as

\begin{equation}
\textbf{I} = \textbf{O} \ast \textbf{P} + \textbf{N},
\label{eq1}
\end{equation}
where \textbf{P} is the PSF of the atmosphere+telescope in the image of interest, while \textbf{O} is the original unperturbed image that one would obtain with a perfect instrument without diffraction and without any atmospheric perturbation. The operator $\ast$ is the standard convolution operator and the quantity \textbf{N} is the noise contribution in the image formation produced at the camera. We assume that we are not in the low illumination regime and \textbf{N} follows a Gaussian distribution with zero mean and diagonal covariance matrix with equal variance $\sigma_{\textbf{N}}^{2}$. The previous expression can be applied to individual monochromatic images, with potentially different PSFs for each wavelength. For simplicity, we make the assumption that the PSF is wavelength-independent, which turns out to be a very good approximation given the wavelength ranges that we are dealing with (less than 2.5~\AA \ in the Hinode/SP case). The specific PSF that we consider is described in \cite{vanNoort2012} and obtained from the pupil specified by \cite{Suematsu2008}, which takes the entrance pupil of the telescope and the presence of a spider into account. Under the presence of uncertainties induced by the noise, any deconvolution must be treated under a statistical framework. Consequently, we only have access to the distribution of reconstructed images. Using the Bayes theorem, the posterior distribution \textit{p}(\textbf{O$\mid$I, P}) which describes the probability of the restored image given the observed image and information about the image-forming system is given by

\begin{equation}
\textit{p}(\textbf{O$\mid$I, P}) =\frac{\textit{p}(\textbf{I$\mid$O, 
P})\textit{p}(\textbf{O})}{\textit{p}(\textbf{I})} ,
\label{eq2}
\end{equation}
where \textit{p}(\textbf{I$\mid$O, P}) is the likelihood or, in other words, the probability that an observed image \textbf{I} has been obtained given an original image \textbf{O} and the PSF. The quantity \textit{p}(\textbf{O}), also named prior, encodes all the a-priori statistical information we have for the original images (i.e., degree of smoothness, presence of large gradients, etc.). This prior contents the a-priori statistical information of the whole field of view of the image. This statistical information could be, for example, that the Stokes $I$ is a function that is defined as strictly positive, or that the magnetic field has spatial correlation between pixels.  Finally, \textit{p}(\textbf{I}) is a normalization constant (termed evidence) that is independent of the unperturbed image. Under the assumption of uncorrelated Gaussian noise in every pixel of the image, the likelihood can be written as

\begin{equation}
\textit{p}(\textbf{I$\mid$O, P})=\prod_{k=1}^{N}{\rm 
exp}\left[-\frac{(I_k-(\textbf{O}\ast\textbf{P})_k)^2}{2 \sigma_{N}^{2}}\right],
\label{eq3}
\end{equation}
where the product is done over all $N$ pixels of the image, $I_k$ represents the $k$-th pixel of the observed image (in lexicographic order), and (\textbf{O}$\ast$\textbf{P}$)_k$ is the k-th pixel (in lexicographic order) of the original image convolved with the PSF. The previous formalism allows us to obtain the maximum a-posteriori (MAP) image, i.e. the image that maximizes the posterior distribution. In addition, we assume that the prior over restored images is flat (all images are equally probable), the MAP solution is equal to the maximum likelihood solution. Assuming that the prior \textit{p}(\textbf{O}) is flat is equivalent to not limit the deconvolution process with any a priori statistical restriction. This solution can be found by taking the derivative of the previous Gaussian likelihood with respect to the original image \textbf{O}. The resulting equation can be solved iteratively with an algorithm known as the Gaussian version of the Richardson-Lucy (RL) algorithm \citep{Richardson1972,Lucy1974}

\begin{equation}
\textbf{O}_{\rm new}=\textbf{O}_{\rm old}+\left[\textbf{I}-\textbf{O}_{\rm 
old}\ast\textbf{P}\right]\otimes \textbf{P},
\label{eq4}
\end{equation}
where the symbol $\otimes$ represents image correlation (which can be written in terms of the convolution operator). The previous iterative scheme does not guarantee positivity of the images even when the observed images are positive and thus somehow have to be forced for Stokes $I$. However, note that for the Stokes parameters, the pixel values can be positive and negative. Additionally, since the RL deconvolution is a maximum-likelihood algorithm, it is sensitive to over reconstruction produced by the presence of noise. The most notable effect is the appearance of high frequency structures in the reconstructed image. To avoid this problem, it is customary to stop the iterative scheme before these artifacts appear.

\subsection{Regularization}

The most straightforward way to deconvolve two-dimensional spectropolarimetric data is to deconvolve the monochromatic images of the Stokes profiles in a way similar to what is done with imaging data \citep[e.g.,][]{vanNoort2005}. This approach presents two drawbacks. First, the number of deconvolutions one has to carry out is large. For instance, the spectropolarimetric data of Hinode SOT/SP contains 112 wavelength points. Second, many of these wavelengths contain practically no information in Stokes $Q$, $U$, and $V$. This is the case of the continuum wavelengths where, unless strong velocity fields are present, the polarimetric signal is expected to be zero. Therefore, one ends up in the difficult situation of having to deconvolve very noisy images. The nature of the RL algorithm then induces an exponential increase of the spatially high frequency noise, making the final images useless.

In general, and as a consequence of the smoothing introduced by the PSF, some information is irremediably lost. This unavoidably transforms the deconvolution process into an ill-posed problem. In particular, a set of solutions with potentially diverging power in the high spatial frequencies are perfectly compatible with the observations. Standard spatial deconvolution techniques solve this dilemma with ad-hoc spatial filtering methods that avoid the divergence of high frequencies during the  deconvolution process. Typical methods include setting a hard or soft threshold on the resulting modulation transfer function that avoids the appearance of high frequencies in the resulting image.

\begin{figure*}
\centering
\includegraphics[width=17cm]{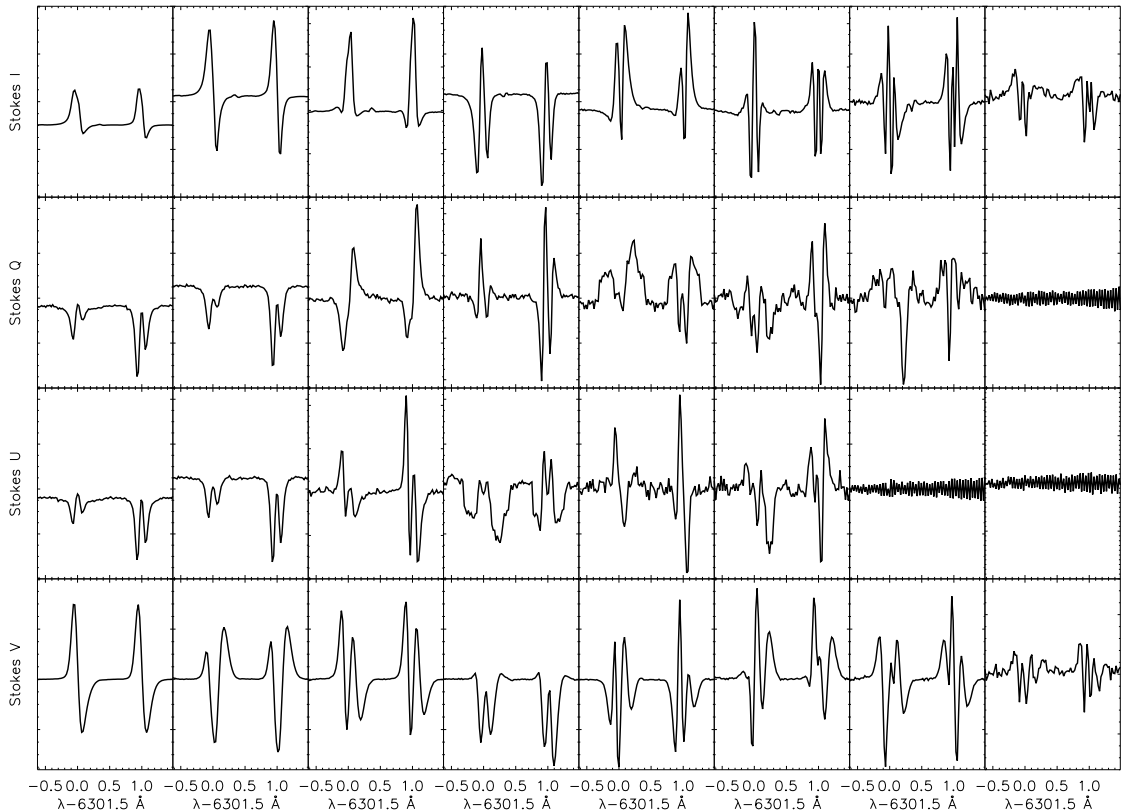}
\caption{First eight eigenvectors obtained after the PCA decomposition of the observed Stokes parameters from Figure \ref{map}. The corresponding eigenvectors for Stokes $I$, $Q$, $U$, and $V$ are displayed from top to bottom and the order of the eigenvectors increases from left to right.}
\label{pca}
\end{figure*}

We pursue a regularized deconvolution. Contrary to the typical procedure in image deconvolution, the regularization that we propose acts on the spectral dimension and not on the spatial dimensions. We assume that the unperturbed Stokes profiles at each pixel can be written as a linear combination of the elements of a complete orthonormal basis formed by the eigenfunctions $\left\lbrace \phi_i (\lambda)\right\rbrace$. Consequently, any of the unperturbed Stokes profiles can be written as

\begin{equation}
\textbf{O}(\lambda)=\sum_{i=1}^{N_{\lambda}} \boldsymbol\omega_i\phi_i (\lambda),
\label{eq5}
\end{equation}
where $N_{\lambda}$ is the number of wavelength points along the spectral dimension. If only a few elements of the eigenbasis are enough to reproduce the unperturbed Stokes profiles, it is advisable to truncate the previous sum and only take the first $N \ll N_{\lambda}$ eigenfunctions into account. Therefore, the unperturbed data is now described by a set of images $\boldsymbol\omega_i$ (that we term projected images), which are built by projecting the Stokes profiles of each pixel on the orthonormal basis functions.

Given that we have assumed that the monochromatic PSF is wavelength independent, the observed perturbed Stokes profiles are obtained after applying Eq. (\ref{eq1})

\begin{equation}
\textbf{I}(\lambda)=\sum_{i=1}^{N_{\lambda}} (\boldsymbol\omega_i \ast \textbf{P})\phi_i (\lambda)+\textbf{N},
\label{eq6}
\end{equation}
where we have used the fact that the convolution operator only acts on the spatial dimensions. Because of the presence of noise, we can find the original unperturbed Stokes profiles by computing the projection of the previous equation onto the orthonormal basis functions

\begin{equation}
\left\langle \textbf{I}(\lambda),\phi_k (\lambda)\right\rangle=\sum_{i=1}^{N_{\lambda}} (\boldsymbol\omega_i \ast \textbf{P}) \left\langle\phi_i (\lambda),\phi_k (\lambda)\right\rangle+\textbf{N} ,
\label{eq7}
\end{equation}
where $\left\langle \cdot,\cdot\right\rangle$ indicates the dot product of the two functions. The noise term still maintains the same statistical properties because the basis is orthonormal, which allows us to simplify the previous expression leading to

\begin{equation}
\left\langle \textbf{I}(\lambda),\phi_k (\lambda)\right\rangle=\boldsymbol\omega_k \ast \textbf{P}+\textbf{N} ,
\label{eq8}
\end{equation}

Consequently, the regularization process we used implies that we have to deconvolve the projected images (associated to the basis functions $\phi_k (\lambda)$) from the PSF and reconstruct the unperturbed image using Eq. (\ref{eq5}). This deconvolution is done using the RL iteration of Eq. (\ref{eq4}).

\begin{figure*}
\centering
\includegraphics[width=15cm]{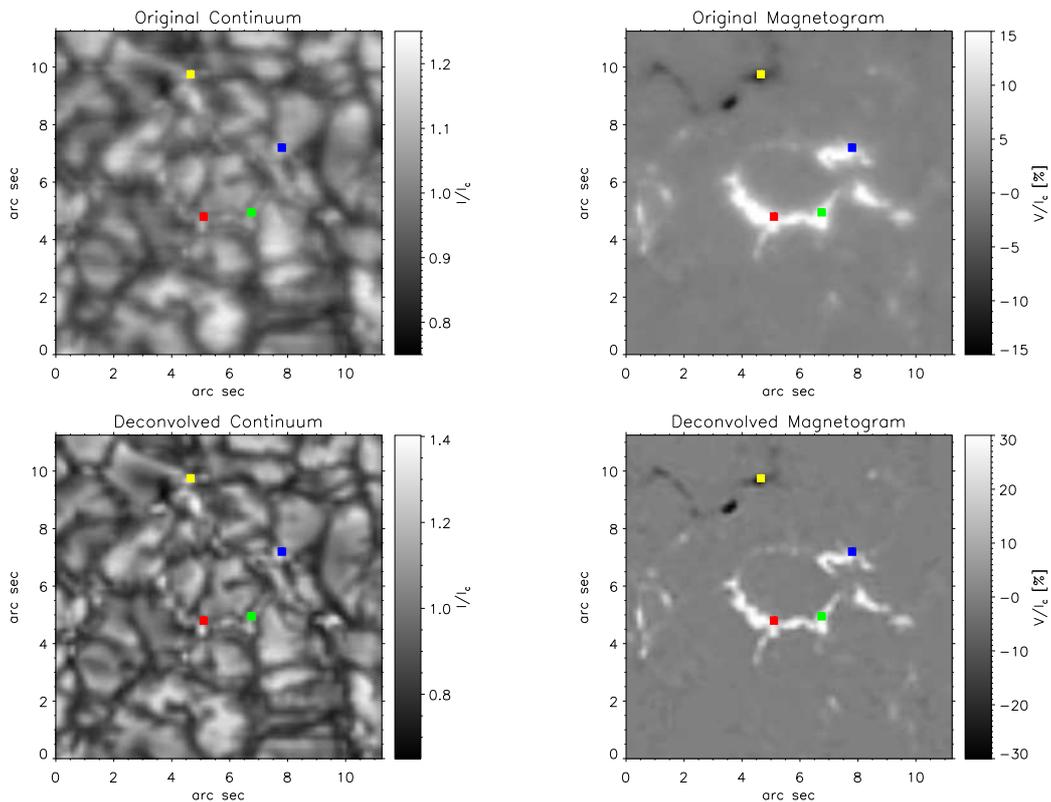}
\caption{Comparison between the observed and deconvolved continuum maps, first column; and the observed and deconvolved Fe~{\sc i} 6302.5 \AA \ magnetograms, second column. The observed region corresponds to the red box in Figure \ref{map}. Finally, the four colored squares indicate the location of the Stokes profiles we examine in detail later.}
\label{snap0}
\end{figure*}

The previous approach is valid for any set of orthonormal functions that one utilizes to explain the Stokes profiles \citep[e.g.,][]{delToro2003}. However, the basis obtained after PCA is ideal in our case because the PCA decomposition transformation is defined so that the first principal component accounts for as much of the variability in the data as possible, and each additional principal component in turn explains the largest variability in the data under the orthogonality constraint. Therefore, working with PCA-projected images, we find that the real signal present in each pixel only appears associated with the first few elements of the basis set, while the remaining elements are used to explain the noise. Consequently, the influence of noise is largely minimized if we only focus on the maps of low-order coefficients. This is a huge advantage with respect to the wavelength-by-wavelength deconvolution.

The procedure starts by building the $M \times N_{\lambda}$ matrix of measurements, where the Stokes profiles (with the mean Stokes profile subtracted) are placed as the rows of a matrix for each one of the $M$ observed pixels. This matrix (or equivalently its covariance matrix) is diagonalized using the singular value decomposition \citep[e.g.,][]{Press1986}. The eigenvectors obtained after the diagonalization form a basis that is efficient in reproducing the observed Stokes profiles and only a few of them are needed. In principle, and according to Eq. (\ref{eq5}), we should use the PCA eigenvectors obtained with the original Stokes profiles. Since we do not have access to those profiles, we have used the observed Stokes profiles to build this database. Unless the original profiles are radically different from those observed, we expect the eigenbasis to be efficient in reproducing the original Stokes profiles also. 

The first eight eigenvectors for the four Stokes profiles computed using all the pixels of Fig. \ref{map} are shown in Fig. \ref{pca}. For the quiet Sun, contrary to the case of an active region, the noise contribution appears in the first PCA eigenvectors (see the 7$^\mathrm{th}$ or 8$^\mathrm{th}$ eigenvector in the linear polarization profiles). This property is a consequence of the predominance of low signal polarization profiles, that need to be carefully taken into account. For our analysis, we only selected the first eight families of eigenvectors for Stokes $I$ and $V$, and the first four families of eigenvectors for Stokes $Q$ and $U$.

\subsection{Comparison with other approaches}

In the case of the inversion strategy presented by \cite{vanNoort2012}, the regularization is done through the selection of the number and position of nodes that describe the physical models. This has the advantage that the physical interpretation of the filtering process is easy: the method eliminates the high frequencies in the Stokes profiles that need perturbations in the depth stratification of the physical magnitudes with more than 3 nodes in depth. Obviously, the number of nodes can be changed at will, but then the inversion of the whole map has to be repeated. 

Our filtering procedure consists, essentially, of a very similar indirect suppression of high frequencies in the image by a filtering of the spectral features. However, since we use an empirical complete basis set that, in principle, reconstructs all the profiles in the field of view to the noise level, we do not eliminate any important spectral feature that is already present in the data. Additionally, given the effective separation between the spatial deconvolution and the non linear inversion of the Stokes profiles, the resulting code is computationally simpler. It has the advantage that any of the existing inversion codes as SIR \citep{RuizCobo1992}, NICOLE \citep{2014arXiv1408.6101S}, Spinor \citep{frutiger00}, or Helix+ \citep{Lagg04} can be used directly. The only addition is a first step in which one has to carry out the spatial deconvolution with a code that we provide for free for the community in the web address indicated in the conclusions.

\begin{figure*}
\centering
\includegraphics[width=18cm]{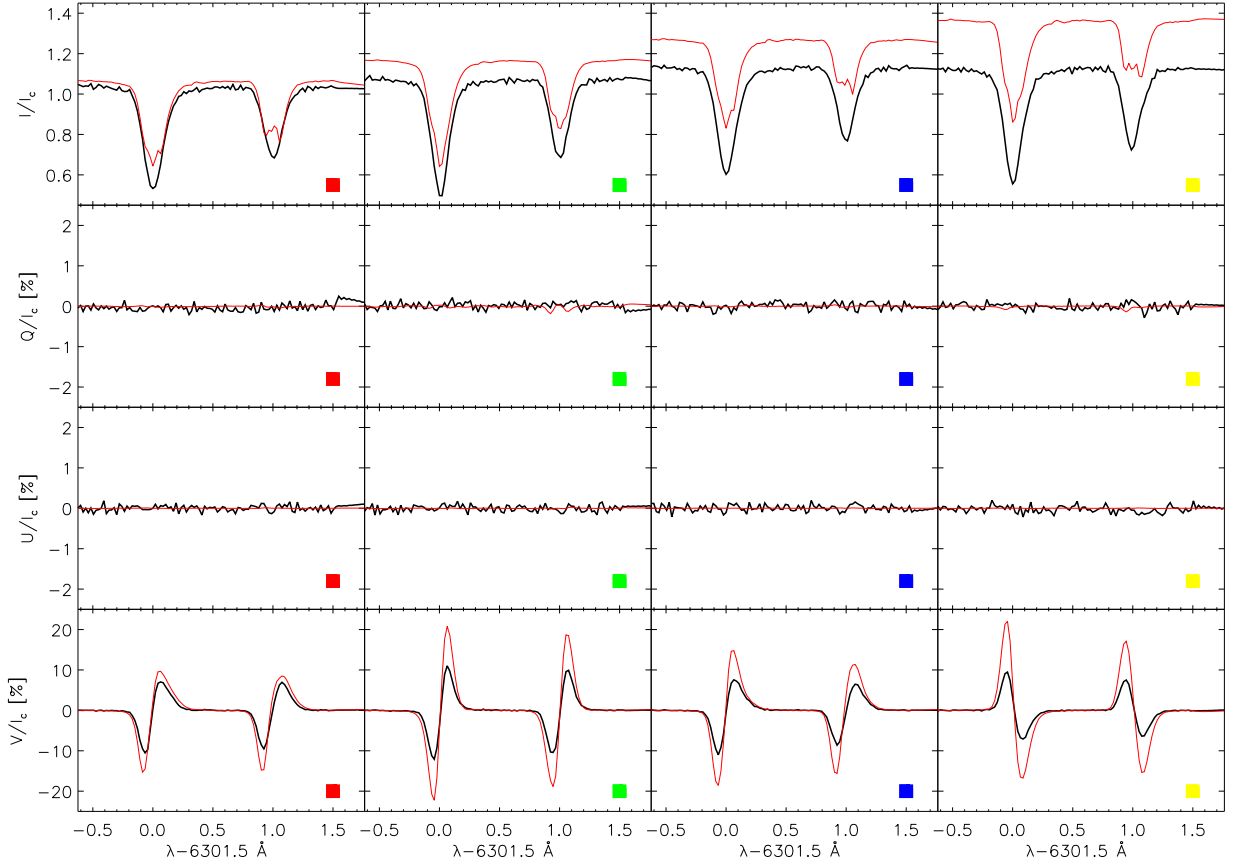}
\caption{From top to bottom, each row corresponds to Stokes $I$, $Q$, $U$, and $V$, respectively. Inside each panel we plotted the original profile in black and the corresponding deconvolved profile in red. In addition, we marked each panel with a colored square to indicate its position on Figure \ref{snap0}.}
\label{per}
\end{figure*}

\section{Properties of the Stokes profiles}

We mainly focus in the analysis of the magnetic patch enclosure inside the red box of Fig. \ref{map}. This magnetic element displays the strongest polarization signals and it is one of the largest structures of the map. It shows an almost circular shape of longitudinal fields that surrounds some granules.

\subsection{Spatial properties}

The first property we study is the changes the deconvolution process induced in the continuum and in the magnetogram images. We  plotted, in the first row of Fig. \ref{snap0}, the original continuum map and the Fe~{\sc i} 6302.5 \AA \ magnetogram (calculated as the difference between the Stokes $V$ profiles at  $\pm$86 m\AA \ from the rest center), while in the second row we show the same magnitudes after the deconvolution process. We see a clear enhancement in the contrast of the continuum features. The contrast, defined as the standard deviation of the brightness in the continuum normalized to its average value, increases from 7.6\% in the original map to 11.9\% in the deconvolved map. The other noticeable difference is the increase in the sharpness of the structure, together with a slight decrease in its size. This effect is probably better observed in the magnetogram, where the structure becomes smaller in the deconvolved data, indicating that the smearing produced by the stray light contamination affects the polarization Stokes profiles. In addition, the deconvolution process also reveals an enhancement in the Stokes $V$ signals in the interiors of the magnetic element. This enhancement indicates that the deconvolved Stokes $V$ amplitude sometimes reaches a factor of 2 larger than its original amplitude. 

Finally, we observed that a reversal in the Stokes $V$ polarity appears in the edges of some structures (e.g., see the small isolated black patch close to the yellow square at coordinates 3.$\arcsec$5, 8.$\arcsec$7 in Fig. \ref{snap0}). These ringing structures may be similar to those reported by \cite{Buehler2015} where they found a magnetic field reversal related to magnetic patches of opposite polarity and magnetic field intensities below 300 G. However, we believe that, in our case, the ringing effects might be generated by the deconvolution procedure itself. We are currently studying spatial regularization techniques (like total variation regularizations) to efficiently suppress these effects.

\subsection{Stokes profiles}\label{Profiles}

The Stokes profiles found inside the magnetic elements with strong longitudinal fields (see white and black patches in Fig. \ref{snap0}) are almost identical in the different regions. The Stokes $V$ profiles are nearly antisymmetric with high amplitudes, between 6-15\% (normalized to $I_{c}$), in the interior of each patch, and their amplitude decreases as we move toward the edge of the magnetic patch. Linear polarization signals, on the contrary, are negligible in these magnetic elements. Finally, Stokes $I$ profiles display a slightly asymmetric red wing that indicates the presence of velocity gradients along the line of sight (LOS) in the center of the magnetic element. 

\begin{figure*}
\centering
\includegraphics[width=15cm]{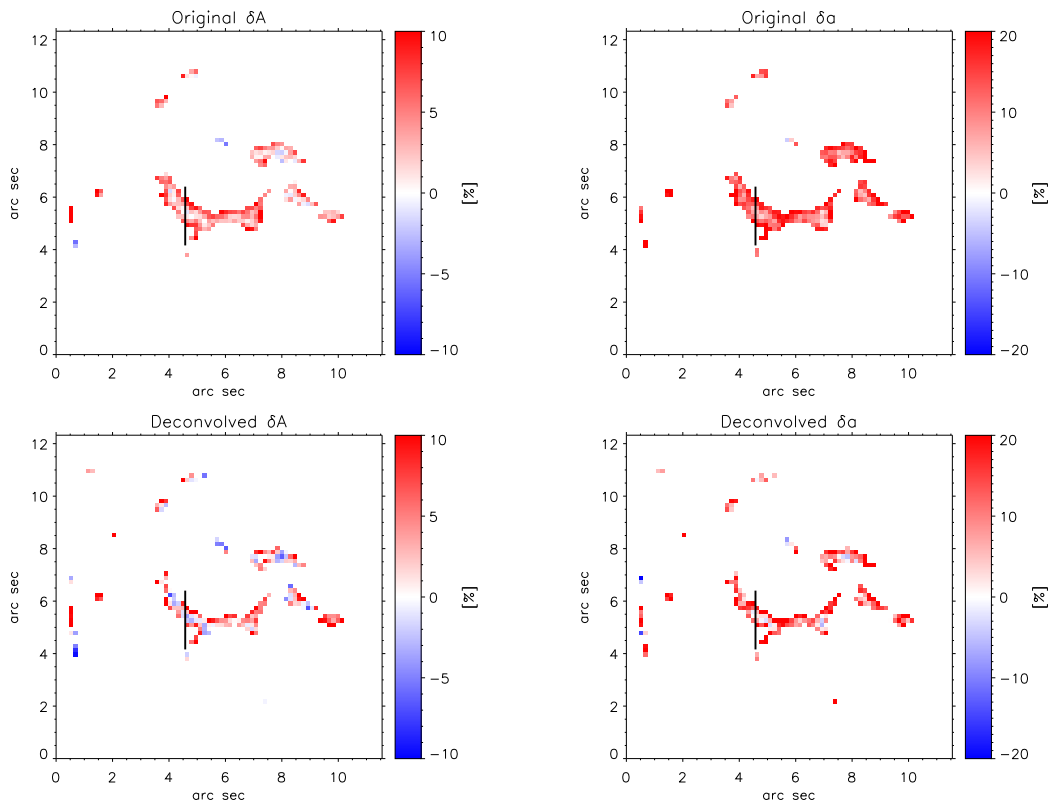}
\caption{Comparison between the original and deconvolved area asymmetry, first column; and the original and deconvolved amplitude asymmetry, second column. The vertical solid line marks the position of the inverted pixels we analyze in following sections.}
\label{asym}
\end{figure*}

To show the effect of the deconvolution process in the Stokes profiles, we have selected four different pixels that display strong Stokes $V$ signals. The position of these pixels are indicated with colored squares in Fig. \ref{snap0}. The corresponding original (black) and deconvolved (red) Stokes profiles are shown in Fig. \ref{per}. 

The Stokes $I$ profile displays major changes, typically presenting a large increase (for granules) or decrease (for intergranules) of the continuum signal with respect to the original signal. Most important, the deconvolved profiles show strong changes in the core of the line. In fact, we detect the effect of the magnetic field producing the splitting of the $\sigma$-components. We believe that this effect is real (and not an artifact of the deconvolution) and related to the presence of a magnetic field because in most cases the splitting is present in the Fe~{\sc i} 6302.5 \AA, which is the most sensitive line to the magnetic field, while it is barely visible in the Fe~{\sc i} 6301.5 \AA \ line, which is the least sensitive to the magnetic field. 

Concerning the linear polarization signals, they are always below the noise level both in the original and in the deconvolved profiles. Circular polarization signals, on the contrary, display strong Stokes $V$ amplitudes, reaching more than 10\% of $I_{c}$ in the original profiles. The deconvolution process affects the Stokes $V$ profiles in, at least, two different ways: it slightly changes the Stokes area asymmetries and it abruptly changes their profile amplitude. As we mentioned before, we find cases where, depending on the surroundings of each pixel, the change of amplitude could reach up to twice of its original amplitude signal.

\subsection{Analysis of Stokes profiles asymmetries}

Area and sometimes amplitude asymmetries are related to the correlation between velocity and magnetic field gradients along the line of sight \citep{Illing1975}. From the results of previous sections, it is clear that the Stokes profiles change after the deconvolution process. For this reason, we intend to study how the Stokes $V$ area and amplitude asymmetries change. To analyze the Stokes $V$ area and amplitude asymmetries, we follow the definition used in \cite{MartinezPillet1997}, that is, the area asymmetry is obtained as

\begin{equation}
\delta A=s\frac{\sum_{i} V(\lambda_i)}{\sum_{i}\mid V(\lambda_i) \mid} ,
\end{equation}
where the sum is extended along the wavelength axis and $s$ is the sign of the Stokes $V$ blue lobe (chosen as $+1$ if the blue lobe is positive and $-1$ if the blue lobe is negative). We choose the range of integration of the Stokes $V$ signals from $-0.43$ \AA\ to 0.43 \AA\ around the Fe~{\sc i} 6302.5 \AA\ line center. 

Likewise, the amplitude asymmetry is defined as
\begin{equation}
\delta a=\frac{a_b-a_r}{a_b+a_r} ,
\end{equation}
where $a_b$ and $a_r$ are the unsigned maximum value of the blue and red lobe of Stokes $V$.

We calculate these quantities over the small fragment of the map contained the red square in Fig. \ref{map}, where we only considered pixels that show Stokes $V$ amplitudes higher than 1$\%$ of $I_{c}$. The results of this study are included in Fig. \ref{asym}. The first row corresponds to the original data set while the second row shows the deconvolved data. The top left panel shows that the area asymmetry is positive for the whole magnetic structure, with maximum values close to 10\%. This area asymmetry is higher at the edges of the structure and decreases to being roughly compatible with zero when we move to the  center of the structure. This effect is also visible in the amplitude asymmetry of the original data (top right panel), which presents values up to 20\% at the edges. The sign of the amplitude asymmetry is also unaltered for the whole structure. 

The maximum value of the area and amplitude asymmetries do not appreciably change in the deconvolved data (bottom row). However, the magnetic element shows some pixels that have changed the sign of both asymmetries, mainly in the center of the structure and around some edges. Similar changes have been found in \cite{Asensio2012} where the authors concluded that the  smearing produced by PSF induces changes in the Stokes $V$ asymmetries which can be partially recovered using reconstruction techniques based on the phase-diversity procedure. This change of sign, especially that of the area asymmetry, indicates that the deconvolution process allows us to detect a change in the gradient of velocity and magnetic field along the line of sight.  In the following section, we study how this information is interpreted by the inversion process to define some relevant properties of the magnetic concentration.

\begin{figure*}
\centering
\includegraphics[width=17cm]{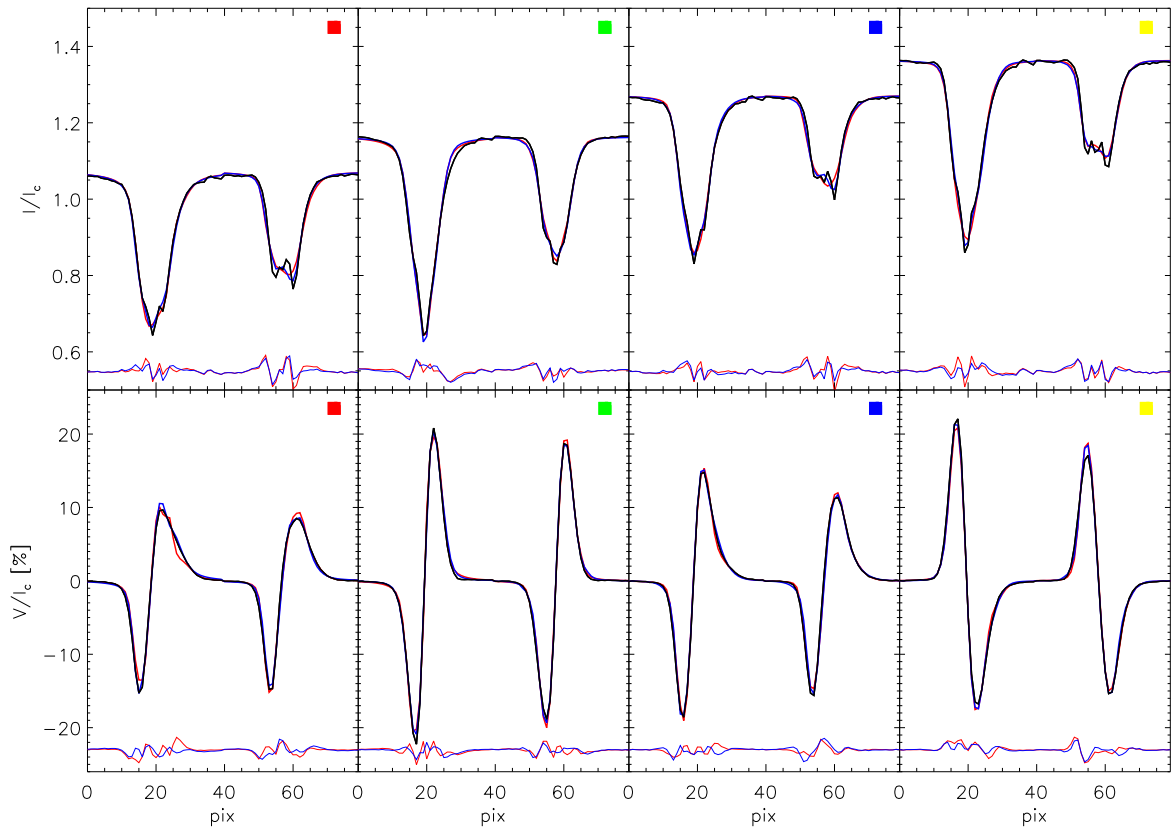}
\caption{Results from the inversion of the deconvolved Stokes profiles presented in Fig. \ref{per} (see the red profiles). We plot the Stokes $I$ profiles in the first row and the Stokes $V$ profiles in the second row. We omit the linear polarization profiles because their signals are always below the noise level. We plot the deconvolved profiles in black, the results from the inversion using the first configuration in red, and the results from the second configuration in blue. We also show in each panel the difference between the deconvolved and the inverted profile using the same color code of the corresponding configuration, and a colored square indicating the position of the pixel on the map of Fig. \ref{snap0}.}
\label{inverper}
\end{figure*}

\section{Stokes profiles inversions}

To obtain physical information of the atmospheric parameters where the Fe~{\sc i} lines form, we carry out the inversion of the Stokes profiles using the SIR \citep[Stokes Inversion based on Response functions;][]{RuizCobo1992} code, which allows us to infer the optical-depth dependence of these atmospheric parameters at each pixel independently.

We analyze the magnetic element shown in Fig. \ref{snap0} with two different approaches, which we describe in detail in the following. The main difference resides that, in one case, we set the microturbulence fixed and equal to zero while, in the other case, the microturbulence was set as a free parameter. The microturbulence correction has been applied in low spatial resolution observations to reproduce the properties of the Stokes $I$ line core \citep[for instance,][]{Westendorp2001}. Although we are using Hinode/SP data, we were not sure if the spatial resolution of these observations is high enough to avoid the use of the microturbulence as a free parameter in the inversions. Thus, we aimed to analyze the results of the inversions using these two configurations, that is, with and without, microturbulence.

\subsection{Solution 1. No microturbulence}\label{conf1}

We employed a single magnetic component parameterized by seven nodes in temperature T($\tau_{500}$)\footnote{The parameter $\tau_{500}$ refers to the optical depth evaluated at a wavelength where there are no spectral lines (continuum). In our case this wavelength is 500 nm.}, five in the LOS component of the velocity v$_{\rm LOS}$($\tau_{500}$), five in the magnetic intensity B($\tau_{500}$), three for the inclination of the magnetic field with respect to the LOS $\gamma$($\tau_{500}$), and one for the azimuthal angle of the magnetic field in the plane perpendicular to the LOS $\phi$($\tau_{500}$). Variables such as micro- and macro-turbulence are fixed to zero and not inverted. At each iteration, the synthetic profiles are convolved with the spectral transmission profile of Hinode/SP \citep{Lites2013}. Since each node corresponds to a free parameter during the inversion, our model includes a total of 21 free parameters.

The number of nodes in B($\tau_{500}$), $\gamma$($\tau_{500}$), and v$_{\rm LOS}$($\tau_{500}$) are necessary to reproduce the small area asymmetries of the observed (see Fig. \ref{asym}) circular polarization profiles \citep{Landolfi1996}, but mainly to reproduce the complex shape displayed by the deconvolved Stokes $I$ profiles. Given that the inferred physical parameters could be reliant on the initial atmosphere, we minimize this effect by inverting each individual pixel with 100 different initial atmospheric models. These initial models were constructed by randomly perturbing the temperature stratification of the Harvard-Smithsonian Reference Atmosphere (HSRA) model \citep{Gingerich1971}. The rest of the physical parameters of the initial model (B($\tau_{500}$), $\gamma$($\tau_{500}$) and v$_{LOS}$($\tau_{500}$)) are extracted from uniform probability distributions and considered to be independent of the optical depth. Out of the 100 solutions obtained for a given pixel, we keep the one that yields the best fit.

\begin{figure*}
\centering
\includegraphics[width=18cm]{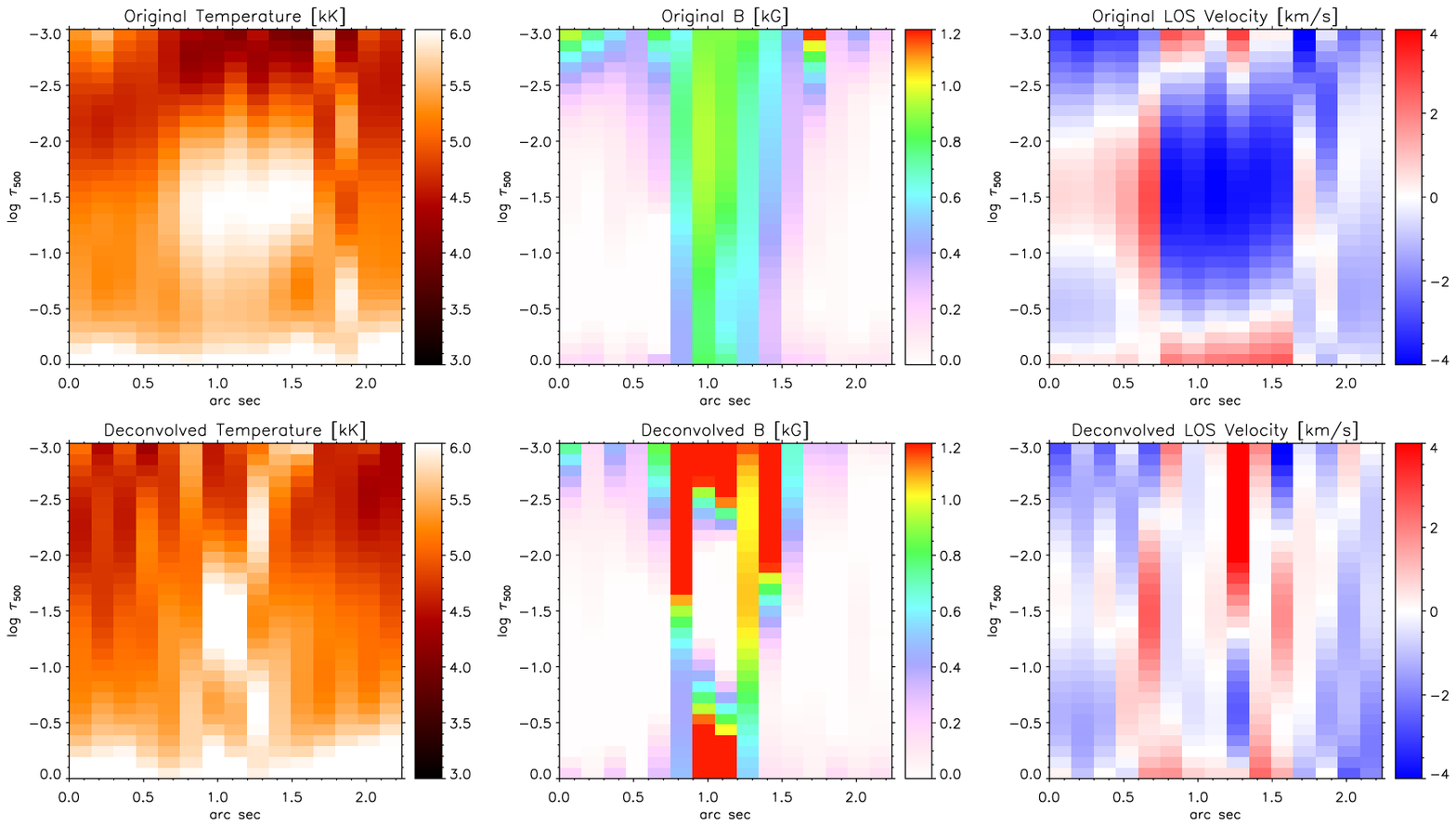}
\caption{From left to right, temperature, magnetic field, and LOS velocity. The color code for the latter indicates in blue the upflowing material while the downflowing material is marked in red. The horizontal axis corresponds to the length of the vertical solid line in Figure \ref{asym} and the vertical axis corresponds to the optical depth.}
\label{s1}
\end{figure*}

\subsection{Solution 2. Non zero microturbulence}\label{conf2}

The second configuration is similar to the first, but in this case we also set the microturbulence as a free parameter. We choose three nodes for the microturbulence, while the number of nodes for the rest of atmospheric parameters remains the same. Then, the total number of free parameters for this second configuration is 24. We also follow the same strategy of inverting each pixel with 100 random inversions and choose the one with the smallest $\chi^2$ value (see Eq. \ref{chi_val}) from the different solutions.

\subsection{Examples}

Some examples of the results of the inversion of Stokes profiles using these two configurations are plotted in Fig. \ref{inverper}. In this figure, we only focus on the deconvolved profiles, although we also inverted the original profiles to compare the results in the following sections. We can see that the resulting fits from the two different configurations (red and blue lines) are close to the deconvolved profiles (black lines). However, although the Stokes profiles are well fitted in general, the core of Stokes $I$ is not well reproduced for the Fe~{\sc i} 6302.5 \AA\ line. We believe that a configuration with the possibility to introduce in the atmospheric parameters abrupt changes in height is necessary to fit these profiles. None of the solutions can be discarded from the point of standard model comparison. However, the solutions using these two configurations have some physical contradictions that we will explain later.

\begin{figure*}
\centering
\includegraphics[width=18cm]{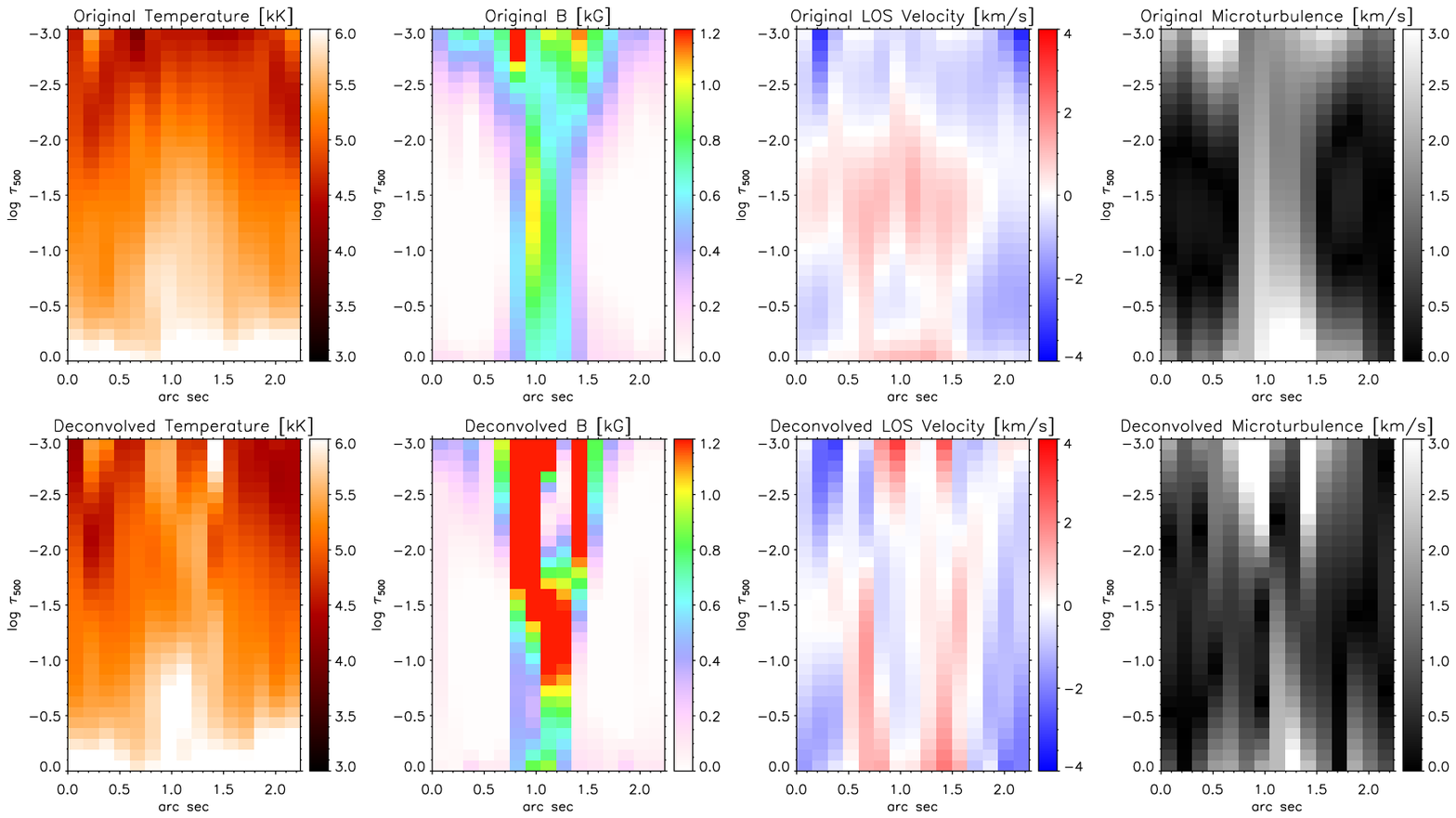}
\caption{Same as Figure \ref{s1}, except we added the microturbulence results at the rightmost panel.}
\label{s2}
\end{figure*}

\subsection{Vertical cut}\label{ver}

The results of the inversion of the pixels marked by the vertical solid line in Fig. \ref{asym} using the first configuration, with no microturbulence, are shown in Fig. \ref{s1}. From left to right, we show the depth stratifications of the temperature, LOS velocity, and magnetic field. The first row shows the atmospheric parameters retrieved form the inversion of the original profiles, while the bottom row shows the same results using the deconvolved Stokes profiles. Concerning the temperature, we find a strong enhancement between $\log \tau_{500}=-1$ and $\log \tau_{500}=-2$ inside the magnetized region. This range is precisely the place of maximum sensitivity of the pair of iron lines at 630 nm. Coincident with the increase in temperature we also find strong magnetic fields. Additionally, the height of the atmosphere at which the magnetic field is strong increases when moving apart from the center of the magnetic structure. This can be taken as an indication that the magnetic field lines are starting to fan out. This opening of the magnetic lines is due to a decrease of  the gas pressure outside the magnetic element. Finally, this increase in the magnetic field and temperature is also consistent with a strong upward motion, while the plasma located outside the central part of the magnetic element (at 0.7$\arcsec$ and 1.6$\arcsec$) shows downflow velocities. The values of the upflow velocities reach  close to 4 km s$^{-1}$.

In the case of the deconvolved data, second row of Fig. \ref{s1}, the behavior of the physical parameters is slightly different. We find a temperature enhancement in the same region, although the spatial size of the structure is significantly reduced. This is because the original structure is smeared by the spatial PSF and the deconvolution process has partially canceled this smearing. The magnetic field structure of the magnetic concentration is now different, showing a strong increase in the center of the structure. These higher intensity values are needed to reproduce the Zeeman splitting displayed by the deconvolved Stokes $I$ profiles and occur at different heights at different pixels. In the central pixels, we find weak magnetic field values at middle heights ($\log \tau_{500}$ between $-1$ and $-2$), and strong values at higher heights. Finally, the velocity of the deconvolved data shows a less smooth solution as compared with the original results.

If we use the second configuration, that is, we set the microturbulence as a free parameter, we find the results shown in Fig. \ref{s2}. The panel distribution is the same used in Fig. \ref{s1}, but we added a column at the rightmost part of the figure that corresponds to the microturbulence results.

Focusing on the original data (first row), the left most panel shows a minor enhancement of the temperature at the center of the magnetic element as compared with the first solution. The magnetic field shows essentially the same structure as Figure \ref{s1}, concentrated field lines along the atmosphere in the central part of the magnetic element, while these lines fan out when we move away from the center part of the structure. The third panel shows major differences between the results of the velocity along the LOS and those obtained with the first configuration. The LOS velocity in the central part of the magnetic element is close to zero and slightly moving downward, while for the rest of the region is slowly moving upward. The changes on the LOS velocity at the central part of the structure are responsible for the inversion of the area asymmetry sign found in the deconvolved data in some pixels, see Figure \ref{asym}. Finally, the microturbulence is large in the central part of the magnetic element and decreases when we move away from the center of the magnetic structure. Its distribution resembles the magnetic field configuration.

The second row of Fig. \ref{s2} shows the results of the inversion of the deconvolved profiles. As occurred with the first configuration, second row of Figure \ref{s1}, the spatial size of the structure is smaller in the deconvolved data case. Likewise, the magnetic field intensity is more intense in the inversion of the deconvolved data. In addition, we also find that some pixels display weak magnetic field values at middle heights ($\log \tau_{500}=[-1,-2]$) and strong magnetic values at higher heights. Finally, the microturbulence also presents similar values between the original (first row) and the deconvolved data (second row), although with the same loss of spatial smoothness between the atmospheric stratification of adjacent pixels as we found for the LOS velocity.

\subsection{The accuracy of the solutions}

It is clear from the previous discussion that two different solutions provide similar good fits. Given that both solutions give different configurations for some physical parameters, it is sensible to compare them using the $\chi^2$ values to see if one of the two solutions is preferable. To do that, we use the following definition of the reduced $\chi^2$:
\begin{equation}
\chi^2=\frac{1}{\nu}\left[\sum_{i=1}^{n_{\lambda}}\left[\frac{I_{i}^\mathrm{obs}-I_{i}^\mathrm{fit}}{\sigma_{I}}\right]^2+\sum_{i=1}^{n_{\lambda}}\left[\frac{Q_{i}^\mathrm{obs}-Q_{i}^\mathrm{fit}}{\sigma_{Q}}\right]^2 \right.
\nonumber \label{chi_val}
\end{equation}
\begin{equation}
\left.+\sum_{i=1}^{n_{\lambda}}\left[\frac{U_{i}^\mathrm{obs}-U_{i}^\mathrm{fit}}{\sigma_{U}}
\right]^2+\sum_{i=1}^{n_{\lambda}}\left[\frac{V_{i}^\mathrm{obs}-V_{i}^\mathrm{fit}}{\sigma_{V}}\right]^2\right],
\end{equation}
where, the superindex ``obs'' and ``fit'' designate the observed profile and the fitted profile. The quantity $\nu$ is the degrees of freedom that corresponds to the difference between the number of wavelength points and the number of free parameters. Finally, $\sigma_{i}^2$ is the noise variance of the original data and it has different values depending on each Stokes profile: $\sigma_I=6.1\times10^{-3}$, $\sigma_Q=7.2\times10^{-4}$, $\sigma_U=7.2\times10^{-4}$, $\sigma_V=7.0\times10^{-4}$. These values are normalized to the mean continuum signal obtained for the whole map.

\begin{figure}
\centering
\includegraphics[width=9cm]{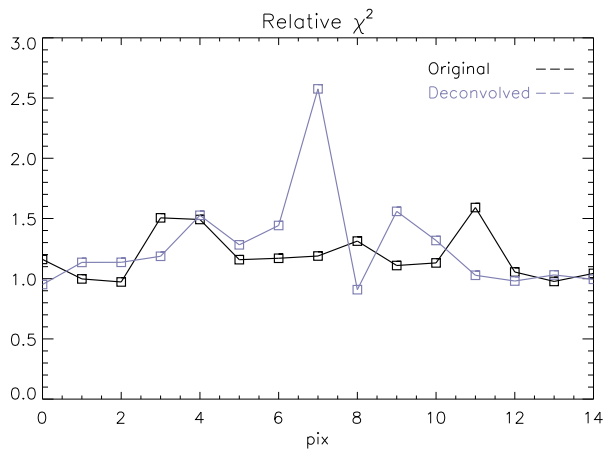}
\caption{ Ratio between the $\chi^2$ values obtained from the second solution, with microturbulence,  over the $\chi^2$ values of the first solution, without microturbulence. Black designates this ratio obtained from the inversion of the original profiles, while gray depicts the ratio between both configurations when we inverted the deconvolved profiles. These Stokes profiles correspond to the pixels marked with the vertical line on Figure \ref{asym}.}
\label{chi}
\end{figure}

We show in Fig. \ref{chi} the ratio between the reduced $\chi^2$ values of the solution with microturbulence over the $\chi^2$ values from the solution without microturbulence. We calculated this value for the inversions of the original and deconvolved data. We can see that they present similar values for the inversion of the original (black line) and the deconvolved (gray line) profiles as both lines are close to 1.  In addition, the standard deviation of the ratio between the reduced $\chi^2$ values for all the inverted pixels is $\sigma=0.4$, and, consequently, both solutions are equivalent because the ratio values are mostly inside 1$\pm\sigma$.  We conclude then that the accuracy of both solutions is similar although the results for the physical parameters are, in many cases, different \citep[see][for a similar conclusion]{2006A&A...456.1159M}. However, we note that the first configuration is favored when using the Bayesian Information Criterion \citep[BIC;][]{schwarz_bic78}, usually applied for approximate model comparison.

\section{Discussion and conclusions}

The analysis strategy (decoupled deconvolution and inversion) we presented is very straightforward to use once the spatial PSF of the instrument is known. Since this is not the case for earth-based observations because of the presence of the fast-changing atmosphere, it is ideal for space-based observatories. We are convinced that the spatial deconvolution of 2D spectropolarimetric data prior to the inversion of the Stokes profiles \citep[or during the inversion,][]{vanNoort2012} is the way to proceed in the near future. The stray-light contamination is correctly treated and one avoids the danger of incurring potential pitfalls \citep[e.g.,][]{Asensio2011}. From the technical point of view, carrying out our analysis is straightforward with the standard tools freely available to any researcher. However, we make our IDL (and Python in the future) code\footnote{http://www.iac.es/proyectos/inversion/deconvolution} available online for everyone to use it, and also for the sake of reproducible research.

The process of spatial deconvolution produces some changes in the Stokes profiles that we studied in detail. The first thing we found is an increase of the continuum contrast of quiet Sun regions from 7.6\% for the original map to 11.9\% in the deconvolved data, accompanied by a reduction of the size of the structures observed at continuum wavelengths and in the magnetogram. In addition, we detect a reversal in the Stokes $V$ polarity at the edges of some of the analyzed magnetic structures. The validity of these polarity reversals has to be studied in future works because it can be an effect of over-reconstruction during the deconvolution process. 

The next noticeable difference that we found is the appearance of the Zeeman splitted $\sigma$-components in the Stokes $I$ profiles in regions of strong longitudinal field. The same pixels also display a sizable increase in the Stokes $V$ amplitude. The analysis of the Stokes $V$ area and amplitude asymmetries revealed that the deconvolution process uncovers negative asymmetries in the central regions of magnetic structures. The absolute value of the maximum asymmetry does not appreciably change.

We examined in detail a vertical cut crossing a magnetic element. After the inversion of the Stokes profiles, we found good fits using two different inversion configurations. The first fit did not include the microturbulence as a free parameter. Using this configuration, we obtained enhanced temperatures on the core of the magnetized region, together with strong upflows. The magnetic field intensity displays, in the original and the deconvolved data, the schematic picture of a magnetic element whose magnetic lines are expanding with height. This scenario can be found in the literature in theoretical works \citep{Grossmann2000} as well as in observational results \citep{MartinezGonzalez2012}.

The second configuration introduces the microturbulence as a free parameter and the LOS velocity distribution completely changes; the plasma inside the magnetic element is slowly downflowing. We also found a temperature enhancement at the locations of the magnetic element, albeit this enhancement is less important. The magnetic structure is similar to the one obtained with the first configuration. Finally, the microturbulence displays low values outside the magnetic structure although it could reaches up to 2 km s$^{-1}$ inside the magnetic element.

Concerning the differences between the inversions of the original and the deconvolved data, we found that the magnetic structure becomes sharper and the smoothness between consecutive pixels is less clear in the deconvolved data. The detected temperature enhancement increases in both cases while the magnetic field intensity strongly increases due to the necessity of reproducing the Zeeman splitting of the Stokes $I$ profiles.

From the two physical solutions obtained in the inversion of the Stokes profiles, we noticed that the velocity configuration of the second solution, see Fig. \ref{s2}, is closer to the results of previous studies related to the solar magnetic elements \citep[see][]{Solanki2006,deWijn2009}. However, the presence of enhanced microturbulence velocity inside the magnetic element is not expected in this part of the structure, which is roughly at rest. In addition, we expect the microturbulence velocity to be more important in outer parts of the magnetic element where we find the interface between the magnetic and non magnetic regions and where abrupt changes on the LOS velocity could be also found. 

It is also interesting to note the lack of coherence of the inferred LOS velocity in the results of the inversion of the deconvolved maps. This makes us think that the inference of enhanced velocities in the central parts of the magnetic structure is partially a consequence of the presence of the PSF. The inversion code fits the broadening of the Stokes $I$ profile with this LOS velocity distribution. The deconvolution process removes this additional broadening and this compensation is not needed.

In spite of using different inversion configurations, we could not perfectly fit the Stokes profiles. In particular, of interest is the Stokes $I$ line core of the Fe~{\sc i} 6302.5 \AA \, that displays the Zeeman splitting but with different intensities in each $\sigma$-component (see Fig. \ref{inverper}). If this is not an artifact of the deconvolution process, this indicates the existence of abrupt line of sight changes of the physical parameters, probably with short height scales, at the line core formation region that could not affect to the Stokes $V$ profiles. These abrupt changes could produce a complex shape of the Stokes $I$, depending on the $\pi$- and $\sigma$-components shapes, while this effect would be not present in the Stokes $V$ profiles. In fact, if the line is forming in this complex environment, the presence of microturbulence on the second configuration of our inversion process is justified.

Finally, we stress the fact that for the first time we have run the deconvolution method presented in \cite{RuizCobo2013} on quiet Sun observations. We have demonstrated that the strong magnetic elements display enough signal-to-noise ratio to reliably reconstruct the information perturbed by the PSF and do not introduce many artifacts. We also point out that our approach allows us to pursue a trial-and-error study (unavoidable when inverting weak signals) that is only possible because we have decoupled the spatial deconvolution and the inversion of the Stokes profiles. Although the true nature of quiet Sun magnetic elements is still far from being completely understood, we conclude that the spatial deconvolution of space solar observation will help to obtain more accurate results.

\begin{acknowledgement}
This work has been partially funded by the Spanish Ministry of Economy and Competitiveness through the Project No. ESP2013-47349-C6-6. AAR also acknowledges financial support through the Ram\'{o}n y Cajal fellowship and the AYA2010--18029 (Solar Magnetism and Astrophysical Spectropolarimetry) and Consolider-Ingenio 2010 CSD2009-00038 projects. \textit{Hinode} is a Japanese mission developed and launched by ISAS/JAXA, with NAOJ as a domestic partner, and NASA and STFC (UK) as international partners. It is operated by these agencies in cooperation with ESA and NSC (Norway).
\end{acknowledgement}

\bibliographystyle{aa} 
\bibliography{network} 

\end{document}